\documentclass[%
  reprint,
  superscriptaddress,
  twocolumn,
  amsmath,
  amssymb,
  10pt,
  aip,
  jcp,
  citeautoscript,
  notitlepage,
  nofootinbib,
]{revtex4-2}

\usepackage[utf8]{inputenc}
\usepackage[T1]{fontenc}
\usepackage{xcolor}
\definecolor{blue}{RGB}{50,50,220}
\usepackage[%
  colorlinks=true,
  allcolors=blue
]{hyperref}
\usepackage{url}
\usepackage{enumitem}
\usepackage{amsfonts}
\usepackage{amssymb}
\usepackage{amsmath}
\usepackage{mathtools}
\usepackage{lmodern}
\usepackage[normalem]{ulem}

\DeclarePairedDelimiter\ppar{(}{)}              % ( )
\DeclarePairedDelimiter\pset{\{}{\}}            % { }

\newcommand{\rref}[1]{Ref.~\citenum{#1}}

\newcommand{\bz}{\mathbf{z}}
\newcommand{\bx}{\mathbf{x}}

\newcommand{\kT}{k_{\mathrm{B}}T}

\begin{document}
\title{Spectral Maps for Learning Reduced Representations of Molecular Systems}
\author{Tuğçe Gökdemir}
\author{Jakub Rydzewski}
\email[]{jr@fizyka.umk.pl}
\affiliation{%
  Institute of Physics,
  Faculty of Physics, Astronomy and Informatics,
  Nicolaus Copernicus University,
  Grudziadzka 5, 87-100 Toru\'n, Poland
}

\begin{abstract}
\href{https://ccp2023.jp/}{Presented at 34th IUPAP Conference on Computational Physics (CCP2023).}
\\[1cm]
Investigating processes in complex molecular systems, which are characterized by many variables, is a crucial problem in computational physics. These systems can be reduced to a few meaningful degrees of freedom known as collective variables (CVs). However, identifying these CVs is a significant challenge, especially for systems with long-lived metastable states. This is because the information about the slow kinetics of rare transitions needs to be encoded in CVs. In this talk, we review recent advances in learning slow CVs and focus mainly on our spectral map technique, a promising deep-learning method that learns CVs based on the slowest timescales. By maximizing the spectral gap between slow and fast eigenvalues of a Markov transition matrix constructed from simulation data, our method effectively captures a simplified representation of alanine dipeptide in solvent. This practical application of our method demonstrates its ability to extract slow CVs, making it a valuable tool for analyzing complex systems.
\end{abstract}
\maketitle

Reconstructing free-energy landscapes of systems as a function of slowly varying reaction coordinates, referred to as collective variables (CVs), is difficult, particularly due to the timescale limitations of molecular dynamics simulations~\cite{valsson2016enhancing}. Many machine learning techniques have been proposed to address this problem. Nevertheless, despite the large success of recent techniques, accurately capturing the slow kinetics of rare transitions between metastable states still poses a challenge~\cite{noe2017collective,wang2020machine,chen2021collective,chen2023chasing}. Among recent deep learning techniques for identifying CVs presented during our talk~\cite{rydzewski2016machine,rydzewski2021multiscale,rydzewski2022reweighted,rydzewski2023selecting}, recently reviewed by us~\cite{rydzewski2023manifold}, is spectral map~\cite{rydzewski2023spectral,rydzewski2024learning}, which we showcase here. As a simple demonstration of our method, we employ spectral map to extract a single CV describing the metastable states and underlying free-energy landscape of alanine dipeptide in solvent.

\begin{figure*}[t]
  \includegraphics{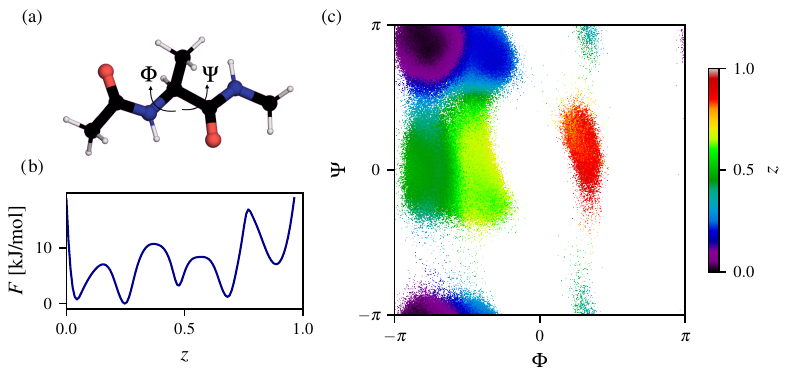}
  \caption{Example application of learning a slow CV using spectral map. (a) Alanine dipeptide in solvent with the $\Phi$ and $\Psi$ dihedral angles shown. As a high-dimensional representation, pairwise distances between heavy atoms are used ($n=45$), calculated from a dataset of three 250 ns trajectories. (b) Free-energy profile along the learned CV $z$ showing five distinct metastable states. (c) Alanine dipeptide conformations in the dihedral angle space are colored according to their CV value.}
  \label{fig:ala}
\end{figure*}

Consider a complex system with high-dimensional representation by $n$ configuration variables (or features), $\bx = (x_1, \dots, x_n)$. To make it easier to understand, we want to create a simplified but physically explainable representation by projecting the system into a low-dimensional manifold, $\bz = (z_1, \dots, z_d)$, defined by a set of $d$ CVs, where $d\ll n$. As CVs depend on the instantaneous of the configuration variables, we can express them as $\bz = {\xi_w}(\bx) \equiv \pset[\big]{ \xi_{k}(\bx), w}_{k=1}^d$, where $\xi$ can be represented by a neural network with parameters denoted by $w$.

To extract the information about the intrinsic timescales of the system whose dynamics is represented in the CV space, we first estimate an anisotropic diffusion kernel~\cite{nadler2006diffusion}: 
\begin{equation}
    \kappa(\bz_k,\bz_l) = \frac{g(\bz_k,\bz_l)}{\sqrt{\rho(\bz_k)\rho(\bz_l)}},
\end{equation}
where $g(\bz_k, \bz_l) = \exp\ppar{-\|\bz_k - \bz_l\|^2/\varepsilon}$ represents the Gaussian kernel, $\rho(\bz_k) = \sum_l g(\bz_k, \bz_l)$ denotes a kernel density estimate (up to normalization), $\|\bz_k - \bz_l\|$ stands for a pairwise distances between CV samples, and a scale constant is $\varepsilon$. Next, we can build a Markov transition matrix by row-normalizing the diffusion kernel:
\begin{equation}
    m_{kl} \sim M(\mathbf{z}_k, \mathbf{z}_l) = \frac{\kappa(\mathbf{z}_k, \mathbf{z}_l)}{\sum_{n} \kappa(\mathbf{z}_k, \mathbf{z}_n)},
\end{equation}
which describes transition probabilities from $z_k$ to $z_l$. Then, we perform a spectral decomposition to estimate eigenvalues of the Markov transition matrix as $\lambda_0 = 1 \geq \lambda_1 \geq \lambda_2 \geq \ldots$, which are used to calculate the spectral gap~\cite{rydzewski2023spectral}:
\begin{equation}
    \sigma = \lambda_{k{-1}} - \lambda_k
\end{equation}
that measures the degree of the timescale separation between the slow and fast variables and $k$ denotes the number of metastable states in the CV space. The main principle behind spectral map is that by using a parametrizable function $\bz=\xi_w(\bx)$, we can adjust $w$ by maximizing the spectral gap, ensuring that the CV space is created by encoding slow degrees of freedom and treating orthogonal fast variables as noise.

Overall, the algorithm for identifying slow CVs using spectral map is composed of the following steps:
\begin{enumerate}
    \item Dataset in a high-dimensional representation is collected from molecular dynamics simulations and used as input. 
    \item CVs $\mathbf{z} = \xi(\mathbf{x})$ expressed by a neural network are trained by maximizing the spectral gap and the degree of separation between slow and fast variables.
    \item Trained CVs are used to build a free-energy landscape, $F(\bz)=-\kT\log P(\bz)$, where $P(\bz)$ is a marginal probability distribution in the CV space, $T$ is the temperature, and $k_{\mathrm{B}}$ is the Boltzmann constant. 
\end{enumerate}

As a simple demonstration, we use spectral map to find slow CVs in a standard testing system for new techniques, alanine dipeptide in solvent (Fig.~\ref{fig:ala}a). As a dataset in a high-dimensional representation, we use three 250-ns molecular dynamics trajectories, carried out at a temperature of 300 K, from the \texttt{mdshare} package. For more details on the dataset, we refer the reader to \rref{wehmeyer2018time}. Distances between the heavy atoms of alanine dipeptide ($n = 45$) are employed as the configuration variables. No preprocessing is performed on the high-dimensional representation. A 5-layer neural network of size [$n$, 50, 20, 10, $d$] is used. ReLU activation functions are applied between each hidden layer. The Adam optimizer is used with a learning rate of 0.0001. The dataset is split into data batches of 1000 samples, and the training is carried out through 100 epochs. Spectral map is trained to extract a single slow CV for $k=5$ metastable states using a scale constant $\varepsilon$ of 0.001.

Our results are shown in Fig.~\ref{fig:ala}. The free-energy profile along the learned slow CV reveals five distinct metastable states with free-energy barriers higher than the thermal energy (Fig.~\ref{fig:ala}b). This indicates that transitions between these states are infrequent and occur on longer timescales. In other words, the slow CV captured by spectral map contains crucial information about the most significant slow processes of the system.

During our presentation, we have delved into the recent advancements of various unsupervised learning techniques for constructing CVs, which are detailed in \rref{rydzewski2023manifold}. However, our focus has been on spectral map, our recent technique. Through a straightforward demonstration, we have illustrated how spectral map can provide insights into the dynamics on longer timescales. By capturing the slowest degrees of freedom, it shows promise in the analysis of molecular simulations of intricate physical systems.

\bibliography{main.bib}

\end{document}